# A Simple Methodology for Model-Driven Business Innovation and Low Code Implementation

Michele Missikoff
IASI-CNR, Rome (Italy)
michele.missikoff@iasi.cnr.it

Abstract: Low Code platforms, according to Gartner Group, represent one of the more disruptive technologies in the development and maintenance of enterprise applications. The key factor is represented by the central involvement of business people and domain expert, with a substantial disintermediation with respect to technical people. In this paper we propose a methodology conceived to support non-technical people in addressing business process innovation and developing enterprise software application. The proposed methodology, called *EasInnova*, is solidly rooted in Model-Driven Engineering and adopts a three staged model of an innovation undertaking. The three stages are: *AsIs* that models the existing business scenario; *Transformation* that consists in the elaboration of the actual innovation; *ToBe* that concerns the modeling of new business scenario. The core of *EasInnova* is represented by a matrix where columns are the three innovation stages and the rows are the three Model-Driven Architecture layers: CIM, PIM, PSM. The cells indicate the steps to be followed in achieving the sought innovation. Finally, the produced models will be transferred onto a BonitaSoft, the Low Code platform selected in our work. The methodology is described by means of a simple example in the domain of home food delivery.

## 1 INTRODUCTION

Enterprise innovation consists in the transformation of some business (e.g, its organization, processes, products, services, etc.) from an existing scenario (referred to as *AsIs*) to a new innovative scenario (referred to as *ToBe*). In enterprise innovation, process innovation (Markic, 2006) assumes a central position. In fact, if our initial focus is on the innovation of, say, product or organization, then we are forced to innovate the involved processes as well. In turn, when we focus on process innovation we need to consider the organization dimension, including enterprise structure, and roles and skills of the people. Furthermore, process innovation requires the development of a new enterprise application capable of supporting the new business scenario. Unfortunately, we are still struggling with the thorny problem of Business / IT misalignment (Luftman, 2000)], i.e., the difficulty for enterprise applications to meet all the business requirements. Model-Driven Engineering (Kent, 2002) recently is straying from the software engineering area to address enterprise engineering. Another important factor is represented by Low Code platforms (Polder, 2019) capable of transforming enterprise models into a running application.

The two mentioned factors are deeply changing the picture, allowing for a strong empowerment of business people who are assuming a leading role in enterprise innovation, including the development of the needed software applications, reducing the role of technical people, in particular software programmers. The advantages of this evolution are enormous, from the reduced time to market of new solutions to the minimal time required for updates and new release of software applications, guaranteed by the disintermediation from technical people in application development. Last but not least, the Model-Driven Engineering and Low Code technology (Gartner, 2020) allow for the progressive solution of the crucial problem of Business/IT misalignment (Aversano, 2012).

In this paper, we propose a methodology, referred to as *EasInnova*, that is grounded in the two above technologies. It has been conceived to allow business people to quickly proceed, in autonomous way or with a minor support from technical expert, in a business process innovation undertaking. MDE provides an effective approach for enterprise modeling, then Low Code platforms offer the possibility of easily move from the business models to a running enterprise application.

Table I - a synoptic view of the EasyInnova innovation method

|     | AsIs | Transformation | ToBe |
| --- | --- | --- | --- |
| CIM | Preliminary models of the existing scenario, problem analysis | Motivations, preliminary solution, pros & cons, selection of candidate solution. Identification of all involved actors. | Preliminary models of the new scenario. Validation with all involved actors |
| PIM | Detailed CIM AsIs models, BPMN diagram | Procedural feasibility. Identify new organization, roles and skills. Progressive building of the ToBe models | New CD, UCD, BPMN. Validation with all involved actors |
| PSM | Analysis of database organization and the critical user functions | Identify: the platform to be adopted; data migration strategy; how to achieve the organizational transformation. | Transfer PIM ToBe models to the selected Low Code platform; generate the new app; implement the new org |

*EasInnova* is based on four pillars: (i) a knowledge management approach aimed at a progressive collection of enterprise knowledge; (ii) a structured approach in modeling the collected knowledge according to the Model-Driven Architecture (MDA); (iii) the transformation of models into running software by using a Low Code platform; (iv) the centrality of business people, involved in the modeling tasks. With respect to the enterprise application, the proposed methodology aims to substantially reduce: (a) the development time, (b) the misalignment of software applications and business needs; (c) the difficulty to timely update software applications to match the fast evolution of business in a globalised market.

In this paper we concentrate on process innovation since it represents the central focus of the majority of business innovations. In fact, even if the primary goal is different, e.g., service or product innovation, it is necessary that the supporting business processes get reengineered in a suitable way. Furthermore, process innovation, as opposed, e.g., to product innovation, is largely independent from the industrial sector, and therefore *EasInnova* is expected to have a wide possibility of adoption.

## 1.1 A Three Stages Scheme of Innovation

When starting an innovation project, we need to consider three stages. The **AsIS** stage where we need to study the existing configuration of the business we intend to innovate. The **ToBe** stage that represents the final scenario, i.e., the new business configuration that we want to achieve. In between we have the **Transformation** stage, representing a bridge between the two. Here the main goal is to identify the final target of the innovation undertaking, but also the strategy to move from the AsIs to the ToBe scenarios. However, a deep understanding of the AsIs scenario, even if we intend to dismiss it, is very important. In fact, in tracing the path towards the ToBe scenario, it is very important to have a clear vision of the starting point. Then, the main effort is on the identification of the ToBe scenario and the transformation process that we need to undertake to achieve it. An early definition of the ToBe scenario is necessarily fuzzy, maybe there are different solutions and various strategies to reach them. Different solutions need to be identified, specified, elaborated, confronted, assessed, to reach a candidate ToBe solution. The whole process needs to be developed with a progressive creation of a body of knowledge, in the form of a collection of models.

We summarise the three stages and their sequencing in Figure 1.

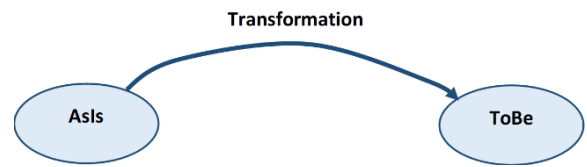

Figure 1: the three stages of innovation

The rest of the paper is organised as follow. In the next sesction, the main lines of the methodology will be presented, introducing the three layers of MDA models: Computation Independent Models (CIM), Platform Independent Models (PIM), and Platform Specific Models (PSM). Then, in Section 3, we introduce the running example and we illustrate how process innovation is addressed in the CIM Layer. In

Section 4 we further elaborate on the models necessary for our purpose and, finally, in Section 5 we address the last PSM Layer where we will produce models ready to be deployed on a Low Code platform. The paper is concluded by Section 6 where some final considerations are reported.

## 2 A MODEL DRIVEN ENGINEERING APPROACH TO BUSINESS INNOVATION

Model Driven Engineering, initially developed in the software engineering domain, is revised in this paper to use it in business process innovation. As anticipated, modelling is the main activity in an innovation project. In particular, we need to collect knowledge about the business process and its context and model it according to the three above indicated stages. Business people and domain experts know better than any other person (techies, in particular) the enterprise they operate in.

Therefore, *EasInnova* starts proposing a simple modelling approach that can be easily adopted by managers and business experts to describe the operational context and the innovation they want to achieve, according to MDA(Mellor, 2002).

MDA organises enterprise modeling according to three layers: Computation Independent Models (CIM), Platform Independent Models (PIM), Platform Specific Models (PSM), aiming to provide a progression in the complexity of the models to be built. They start from simple models in the CIM layer and then provides at each successive layer additional knowledge that, in a coherent manner, builds a path from an informal and incomplete representation to increasing levels of formality and completeness. This progression facilitates the involvement of people without a technical background.

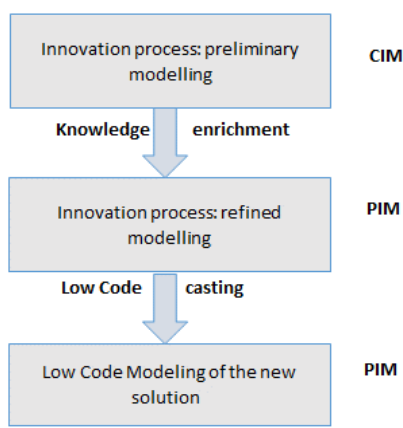

Figure 2: The MDA layers in business process innovation.

Figure 2 illustrates a sketchy view of the three progressive layers: CIM, PIM, PSM, and the kinds of models that MDA requires for each of these layers, in the perspective of business process innovation. Please note that the last level PSM is the final one where, thanks to a Low Code platform, the built models will be easily transformed into a running enterprise application.

The rationale of *EasInnova* is based on the synthesis of the three MDA levels and the three innovation stages introduced in the previous section.

It is important to clarify that in the innovation process the three layers are not achieved in strict sequence. At first, we start modelling the CIM layer, at our best. The PIM models build on top of the CIM models, by adding more knowledge. But this enrichment is used also to validate, and in case to further enrich, the CIM models, with the objective to maintain a global consistency. The same is true when we address the PSM modelling. In the sequel, this spiral approach will be better clarified.

In *EasInnova* the models that represent the enterprise knowledge can assume various forms, with a progression of details and levels of formality. In particular, we have: (i) *plain text*, a narrative form of knowledge representation; (ii) *structured text*, e.g., itemised lists (bullet points, numbered lists, etc.) that collect and organise short statements; (iii) *tables*, typically providing a systematic visualization of knowledge items; (iv) *diagrams*, where the knowledge is graphically represented, according to a given standard. In the final stage of application development, there can be additional forms of knowledge representations: from mathematical expressions to decision tables, from scripting to traditional software coding. However, adopting a Low Code platform, this final part, not relevant for the business problem, will be minimal and will not be addressed in this paper.

In the next sections we elaborate the three MDA layers in the context of the business innovation lifecycle.

### 2.1 The *EasInnova* Methodology

In this section we provide a description of the proposed method for business process innovation. As anticipated, *EasInnova* is centrally based on the three stages: AsIs, Transformation, ToBe, and the three layers of the Model Drive Architecture: CIM, PIM, PSM. Their interbreeding allows us to build a three by three matrix. This is illustrated by the Table 1 where the three stages of the innovation process label the columns and the three layers of MDA label the rows.

The guidelines of *EasInnova* suggest to proceed traversing the Table 1 according to the two axes. Horizontally, from left to right, moving across the three stages of innovation: AsIs, Transformation, ToBe. Vertically, from top to bottom, proceeding with a progressive knowledge enrichment: from a preliminary knowledge (CIM), to the layer below, focussing on behavioural knowledge (PIM), and then, reaching the lowest layer (PSM), by providing Low Code specific models.

The cells indicate the steps that need to be achieved. Starting from the first step: CIM-AsIs, we proceed with the second: CIM-Transformation, reaching the third and last of the first row: CIM-ToBe. Afterwards, in a similar way, we address the second row, PIM, and then the third and last row, PSM. The end of the innovation process is achieved when we complete the last step: PSM-ToBe. Please note that we described the procedure as a sequence, however, in practice, the process is not linear and each time we proceed, adding new knowledge, we have the opportunity to go back to validate (and possibly enrich) the models produced in the previous steps.

As anticipated, a Low Code platform provides a high level development interface that allows us to transfer the PSM models on the operational platform in order to quickly release a running application. The majority of Low Code platforms offer a dedicated subsystem that allows us to build the PIM knowledge already on the platform. Even if it appears an advantageous feature that may further reduces the development of the application, we prefer to adopt a different modellong environment for the first two layers (in our work we adopted *LucidChart®*, see below) that is 'platform agnostic', to avoid any risk that a technology environment may hinder the work of non-technical people, creating a bias towards a specific technology.

In the next section, we adopt a running example to illustrate in practice how the *EasInnova* methodology can be applied. The running example that we use is a home delivery pizza shop, PizzaLove. It accepts orders via Web and delivers pizzas at home.

## 3 The CIM Layer

In *EasInnova*, the CIM layer is elaborated for each of the three stages of the innovation project, therefore we will have the following steps: CIM-AsIs, CIM-Transformation, CIM-ToBe.

### 3.1 CIM-AsIs Step

In this step, the models to be built start with a preliminary plain text description of the business scenario, and specifically the business process to be innovated. The text below is the synthesis of a (fictitious) interview of the manager of PizzaLove.

**AsIs Business Scenario**
*PizzaLove is a home delivery pizza shop. The customer fills in the order and then submits it to the shop, with the payment, by using a Web portal. Making good pizzas requires good quality dough, produced in-house, and a careful baking of the pizza. To make clients happy, we need to quickly fulfil the order and the delivery boy needs to know streets and how to speedily reach the customer's address.*

Once we have the synthetic description of the business process, we perform a preliminary semantic analysis of the text according to the OPAAL scheme (Object, Process, Actor, Attribute, Link), elaborated along the line of a business ontology (D'Antonio, 2007). The analysis will proceed lifting the collected knowledge to the conceptual level. In particular, the terms are classified as: (i) *Object*: passive entities with a lifecycle that follows to the CRUDA paradigm, i.e., the traditional Create, Read, Update, Delete (Martin, 1983 and Torim, 2912), to which we add Archive that is particularly relevant in business processes; (ii) *Process:* a sequence of tasks aimed to enact CRUDA operations on one or more business objects; (iii) *Actor:* is an active entity involved in one or more processes;

Table 2 – AsIs OPAAL Lexicon

| Categories | Business terminology |
|---|---|
| *Objects* | Order, Pizza, Dough, Home, Address, … |
| *Processes* | Backing, MakingDough, SubmittingOrder, ReceivingOrder, DeliveringPizzas, … |
| *Actors* | PizzaShop, Customer, DeliveryBoy, … |
| *Attribute* | Address, Price, Quantity, PizzaKind, … |
| *Link*s | Customer-Order, Order-Pizza, Home-Address, Pizza-Backing, Pizza-Deliverying, DeliveryBoy-Delivering, Customer-Address, Pizza-Pices, Order-Qty, … |

(iv) *Attribute*: a property associated to one of the former concepts; *Link:* a relationship between two of the above entities. Table 2 reports the OPAAL

Lexicon (it may evolve into a Glossary when we add the description of the terms, However, for sake of space, we omit this richer structure).

Please note that in the table the terms labelled as Process are represented with the gerund to emphasise that they represent verbs (terms like Order may cause confusion).

Starting from the descriptive text and the OPAAL Lexicon, we can build a first UML-Class Diagram (CD) [ref] by using the terms collected in the Lexicon. In the class diagram, boxes are labelled with one of the identified terms, and attributes are reported within the box. Pairs of terms in the Link section are represented by arrows, in general labelled with the actions reported in the Process section (Please note that the term Process in OPAL is more general than 'business process', including various notions, such as task, operation, action, activity, function). All class and arrow names need to be already identified and reported in the OPAAL Glossary. In certain cases, a link connects a complex attribute (e.g., Address). In this paper we omit the Class Diagram, since we are focussing on business processes.

Please note that at this stage we need (possibly) to be precise, but we don't need to be neither formal nor complete. In fact, with *EasInnova*, we do not proceed in a linear way, but rather in a spiral way. For instance, when drawing a CD it may be the case that new terms, not yet identified, will emerge, then we go back to the OPAAL Lexicon (Please note that we use the term 'lexicon' and not glossary since this structure collects the sole terminology without the description of terms) adding the new terms to it, in order to keep the different models aligned.

**Problems analysis**

Now we need to explain what are the motivations that are pushing to change the existing business process.

We sketchily classify the motivations into *problems*, when the innovation is needed to correct a situation hampering the business, and *desires*, when we address a possible improvement. It is important to note that in this phase we concentrate on the analysis of the AsIs *status quo*, avoiding any anticipation of possible solutions. The problem analysis can be carried out by means of interviews, questionnaires, and any other means typical of knowledge elicitation (Gavrilova, 2012).

Table 2 reports two problems and a desire that have been.

At the end of this step, we will share the models that have been created, namely: textual description, OPAAL Lexicon, Class Diagram, and Problem Analysis Table 3, with all the actors and stakeholders currently involved with the business that will be transformed.

Table 3 – Problem Analysis

| Problem label | Problem description |
|---|---|
| *Problem 1* | When many orders arrive at the same time, the delivery time gets too long |
| *Problem 2* | If the DeliveryBoy is sick, then it is necessary to move someone from the kitchen to the delivery function |
| *Desire 1* | The shop would like to dedicate a greater attention to the customer care |

'External' opinions, suggestions, and criticisms are important elements to be collected before moving to the next step. They will be carefully considered to refine all the AsIs models.

### 3.2 CIM - Transformation Step

Having analysed the problems, in this step we first formulate an *Innovation statement*, a synthetic textual description of the innovation goals.

Table 4 – Innovation Strategies

| Problem label | Problem description | Innovation strategy |
|---|---|---|
| *Problem 1* | When many orders arrive at the same time, the delivery time gets too long | More staff for pizzas cooking, and efficient partners for delivery |
| *Problem 2* | If DeliveryBoy is sick, then it is necessary to move someone from the kitchen to the delivery function | Partnership with (one or more) Delivery Service(s) will substantially reduce this problem |
| *Desire 1* | The shop would like to dedicate a greater attention to the customer care | The creation of a Customer Relationship section will meet this |

**Innovation statement**
*The current innovation intends to reduce the time needed to fulfil a customer order, in particular avoiding the delay that happens when by too many orders arrive at the same time. The response time depends the dough making and the efficiency of the*

Table 3 – Solution Analysis

| Label | Description | Pros | Cons | Mitigation |
|---|---|---|---|---|
| Solution1 | Strengthen the business by hiring more people | We will dedicate more human resources to the different functions of the shop: customer relationships, dough preparation, pizza making, pizza delivery. | - Higher costs<br>- More human resources to be managed<br>- Risk of overstaffing in low business periods | - Improve efficiency<br>- Improve HR management<br>- Improve smart working and job flexibility |
| Solution2 | Reorganise the business by focusing on core functions (making pizzas and customer care), externalising non-core functions (dough preparation and pizza delivery) | - Increase flexibility to face peak of orders,<br>- Move idle staff to core functions<br>- Increase the volume of fulfilled orders<br>- Improve the quality of customer relationship | - Costs of a deep organizational change<br>- Difficulty in selecting reliable business partners (DoughFactories and DeliveryService) | - Select a suitable change management methodology<br>- Delegate the selection to a specialised agency |

*Delivery Boy, both appear as weak factors in timely fulfilling an order. Finally, we wish to improve the relationships with customers.*

The Innovation statement, together with the problem analysis of the previous step, is further elaborated to identify a possible solution for each specific problem. The outcome of such analysis is reported in Table 3.

Then, we need to further elaborate on possible solutions, aiming at identifying the candidate one. For each solution we list *pros* and *cons*, and possible mitigation strategies to minimise the hindering factors. In Table 4 two different solutions are reported.

For each listed solution, it is necessary to further analyse business actors, highlighting what kind of impact will bring. Such an analysis can be carrid out by means of a dedicated method [ref], such as *SWOT* (Strength, Weakness, Opportunity, Threat) analysis (Hill, 1997) for each solution. Or any other preferred method, such as TOWS (Kulshrestha, 2017) or SOAR (Strength, Opportunities, Aspiration, Results). The details of such an analysis falls outside of the scope of this paper.

**Identification of the candidate solution**

In our example we assume that the Solution 2 is prevailing against Solution 1, therefore the former is selected as the candidate solution. Please note that, according to the non-linear pattern of the innovation process, a solution identified in this step may result inconvenient when addressing the subsequent steps. In this case, we need to go back, enriching the knowledge previously collected in the Solution Analysis table, in case discarding the first candidate solution if favour of another one.

## 3.3 CIM-ToBe Step

Having selected the Solution 2, we follow for the ToBe scenario the same mechanism adopted in the CIM-AsIs step. We start with a short text to better explains the solution sketchily described in the Table 4 and then we provide a first version of the corresponding OPAAL Lexicon, reported in Table 5, and the new UML-CD. We do not report such a diagram that can be easily derived from the Table 5, with a similar approach adopted in the CIM-AsIs step.

**ToBe Business Scenario**

*The new business process requires a new organization based on three business units. (1) Customer Relationship unit that receives from the customer the order with the payment and, after the delivery, collects the appraisal of the customer; (2) PizzaCook, the core business units that prepares and bakes the ordered pizzas; (3) Dough Procurement unit that guaranties the replenishment of the dough stock. Then, there are two strategic partners: (A) Delivery Service that, once the pizzas are ready, is alerted for pickup and delivery to the client. (B) Dough Factory that, when the stock is below the threshold, supplies the needed dough.*

Please note that the terms reported in this Lexicon are derived from the business scenario description, but they may be possibly renamed to make them easier to be used as labels in a technical context.

Table 6 – ToBe OPAAL Lexicon

| Categories | Business terminology |
|---|---|
| Object | Order, Payment, Pizza, Dough, Home, Address, … |
| Process | Backing, MakeDough, SubmitOrder, ReceiveOrder, CookPizzas, CollectPizzas, Delivering, AlertPizzasReady, CustomerPolling, … |
| Actor | CRM, PizzaCook, SCM, Customer, DeliveryService, DoughMaker, … |
| Attribute | Address, DoughThreshold, DMContacts, DSContacts, … |
| Link | Customer-Order, Order-Pizzas, Customer-Address, Pizzas-Cooking, Pizza-Deliverying, DeliveryBoy-Delivering, … |

Please note that Table 6 has mainly a demonstration purpose and is not complete.

Having terminated the elaboration of the three steps of the CIM layer, we proceed addressing the PIM layer that represents the core of the innovation undertaking.

## 4. THE PIM LAYER

This layer is primarily concerned with the behavioural dimension of the knowledge, here we need to fully specify and precisely model the business processes, both for the AsIs and the ToBe scenarios.

Here the knowledge is primarily represented by diagrammatic models, to this end we propose an international, universally accepted standard: BPMN (Business Process Modeling and Notation, proposed by OMG) [ref.]. Then, besides the BPMN diagrams, the PIM layer requires the modeling of actors, objects, and business documents. In particular, the user interactions are modelled by using the UML-UCD (Use Case Diagram) [ref] and a detailed UML-CD is built by enriching the simplified Class Diagram produced earlier. The content of the two diagrams needs to be consistent with the ToBe OPAAL Lexicon, in particular UCD can be derived from its Link section. As anticipated, when elaborating the models of the PIM layer we need to check their coherence with the CIM models and, in case, go back to update the latter.

The full understanding of the AsIs scenario is a key issue, it is risky to start an innovation undertaking without a correct understanding of the existing reality. Luckily, a full specification is possible for the AsIs scenario, however things get more complex when we consider the ToBe scenario, our target, and the Transformation. It is an inherent characteristics of an innovation venture to start from a partial definition of the ToBe scenario. But also the transformation process, the strategy to evolve from the AsIs to the ToBe scenario, cannot be fully defined when we start.

Following the same organization of the CIM layer, in the first step we start modeling the existing AsIs business process, while in the ToBe step we model the target scenario, progressively defined in the Transition stage.

In summary, at the end of the modelling effort, in the PIM layer, we will have 3 kinds of diagrams for each scenario: BPMN, to represent the behaviour of the actors, UML-UCD to represent the user interaction with the application, UML-CD, to model (in greater detail than the CIM layer) the entities and their relationships. In this paper we concentrate on the business processes, omitting the other two diagrams.

The BPMN diagrams reported below are sufficiently simplified to result rather intuitive also for a reader who is not familiar with the such a notation.

### 4.1 PIM-AsIs Step

In Figure 3 we report the BPMN model of the existing process. Please note that (according to the BPMN notation) we have two pools that correspond to the two primary actors: *Customer* and *PizzaLove*. To keep the picture simple, we avoid to represent some internal activities and actors, such as the DeliveryBoy or the DoughMaker. Additional details will not contribute to enrich the overall description of the *EasInnova* methodology, conversely it may distract the reader from the essence of the proposal.

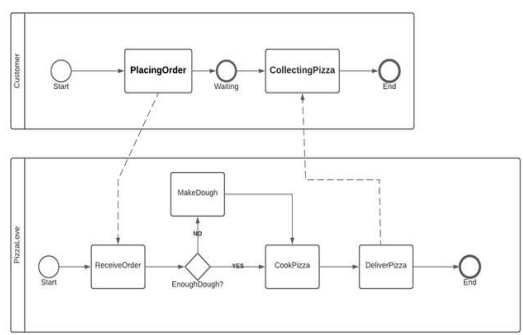

Figure 3: The AsIs Business Process

Figure 3 shows how the pizza shop currently operates, with a rather simple organisation that leave space to many possibilities of optimization. For instance, the idea that we prepare the dough on the fly when we process an order and we discover that the available dough is not enough to fulfil it. Doing so, we may delay the cooking and the delivery of the ordered pizzas. This consideration provides additional information to the Transformation stage.

### 4.2 Transformation and ToBe Modeling in the PIM Layer

PIM-Transformation and PIM-ToBe are the last steps of the innovation venture aimed at producing the new business scenario, in particular the new process models. In fact, the remaining layer, PSM, concerns only with technical aspects.

In particular, the activities of the PIM-Transformation stage, focussed on the definition of the new processes, are based on previous experiences and a few guidelines and practices, summarised below.
- A focus on the core business, by enhancing the staff dedicated to the preparation and cooking of pizzas, and to customer relationships.
- Activities dedicated to make dough and deliver pizzas will be externalised to (carefully selected) partners.
- New business functions will be introduced in the pizza shop:
  o CRM: Customer Relationship Management, to interact with the customers, receiving their orders and collecting their opinions after the service. Furthermore, the CRM elaborates the collected data for a constant monitoring and the improvements that can be introduced to meet customer needs.
  o SCM: Supply Chain Management, to guarantee the provisioning of dough from more than one partner (useful in case of a peak of demand). In parallel, a dough stock, periodically replenished, is managed in a fashion suitable to guarantee fresh dough and to avoid any shortage.

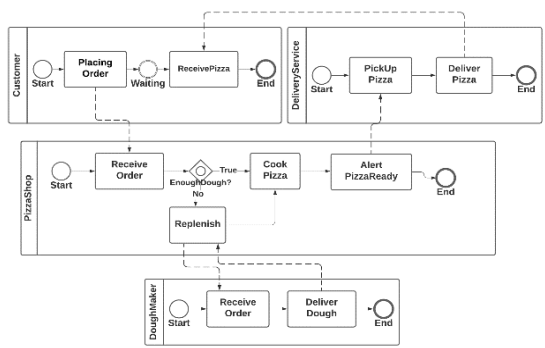

Figure 4: the ToBe Business Process

In the diagram of Figure 4 we have four pools, corresponding to the pizza shop and the two strategic partners, beside the customer pool. Please note that the diagram represents the business process schemes at the conceptual (intentional) level. Therefore, the real actors (instances) in the concrete scenario can be more than one for each pool. For instance, we may have more than one instance of *DoughMaker* and *DeliveryService*. Then, when a process instance is activated by a customer order, the *PizzaLove* shop may decide which partners will be involved (to keep the diagram simple we avoid representing details of this task). Please note also that the replenishment task issues a dough order and then continues, in parallel the DoughMaker process proceeds independently, fulfils the order and then terminates.

# 5 The PSM LAYER

The two previous layers are fully centred on the business perspective, this one is centred on the technology. This is the third and final modeling layer that yields the running application. In earlier times, this modelling activity consisted in additional efforts in terms of technical coding, even if based on a high level language, such BPEL: Business Process Execution Language (van der Aalst, 2005).

Things have substantially changed with the last generation of Low Code platforms that are able to close the gap between the higher level modelling of the business scenario and the lower level of software development activities. The models created in this layer are used to generate a running application, with a minor coding effort.

In fact, instead of proceeding further downward to provide more detailed computational models (e.g., flow-charts as the most 'low level' model, preceding the actual coding), the Low Code technology allows for the lifting the level of (machine) abstraction, being capable of directly executing more abstract models, without (or with a limited) need of software coding.

While the overall architecture of the business application, including actors, objects, and processes, have been fully developed in the previous steps, the implementation platform requires a few additional details. For instance, in the PIM-ToBe step we build a business process indicating the tasks (besides the other BPMN elements) at a logical level. In this layer we need to further enrich such a knowledge, for instance indicating if a given task is manual (i.e., performed by humans) or automatic (i.e., performed by a machine). Furthermore, for the former we need to specify if it's a User Task, therefore requiring a user interaction (implying the definition of a user interface) or if it is a 'pass through' manual task, where humans perform the task without the need to when the task has been correctly completed).

## 5.1 PSM-AsIs and Transformation steps

An important objective of this layer is represented by the data migration operation, to move the business data from the legacy application (AsIs) to the new application (ToBe). This task is achieved in the PSM-AsIs step, for what concerns the analysis of existing data assets, and the PSM-Transformation step, for what consists the construction of a strategy to achieve the porting of business data to the new application.

If we assume that the enterprise is not yet committed to working with a specific Low Code platform, another important objective of the PSM-Transformation step is to analyse the various options offered on the market and to select the right platform.

This can be a very demanding task. In our case we decided to start from a literature analysis concerning some of the most popular Low Code platforms, in particular we considered OutSystem, Mendix, Zoho, Oracle, Camunda, BonitaSoft. Then we created a shortlist with the last two platforms, primarily because they are both free and open source (for the Community edition) and their characteristics and performances are not so distant from their commercial alternatives. Hence, we installed the two platforms carrying out a preliminary trial. Eventually, Camunda was selected, primarily for its smoother learning curve, the quality of available documentation, and its modularity. In the future we plan to carry out a more systematic benchmarking (IT Central, 2020) between Camunda and BonitaSoft to reach a more knowledgeable choice.

### 5.2 PSM-ToBe Step

This is the last step where we finalise the detailed models necessary to produce the running prototype. As anticipated, here the main objective is to represent the business process models, defined at a logical level in the PIM-ToBe step, according to the selected platform. Since this last step produces executable models they need to be precise, unambiguous, and complete with respect to the computation that the different functions and operations require.

Then, the running prototype cannot be achieved with business process modeling alone, then some (minor) quantity of coding is actually required. The quantity of coding much depends on the level of customization required and the software extensions available for the selected platform. On the market there is plenty of plug-ins and add-ons available and Camunda supports their easy inclusion in the application.

In Figure 5 we report the final diagram that has been generated with the Camunda Modeler. It looks very similar to the PIM-ToBe model, in the structure, actors, and task organization, with a few additional symbols (please refer to (Camunda, 2020) for details).

Furthermore, the Camunda Modeller offers a simple simulator that allows us to achieve a first validation of the business process before moving it to the execution engine. In the first prototype, both data management and user interfaces have been preliminarily implemented by using the simple facilities offered by the platform. A final engineering phase will be required to develop a fully-fledged user interface and database management. To this end, Camunda offers a number of solutions that facilitate this work, allowing also business people to participate with an active role.

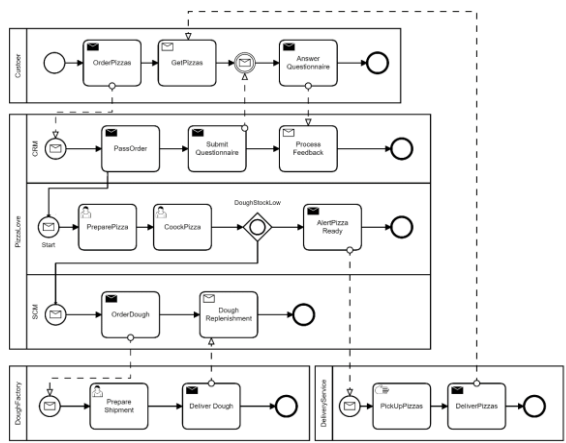

Figure 5: Camunda Modeler Business Process

## 6 CONCLUSIONS

In this paper the main lines of the *EasInnova* methodology have been presented. Such a methodology has been conceived with the objective to empower business, giving them the possibility to assume a central role in enterprise application development. One of the main advantages of the proposed methodology is the possibility of carrying out in parallel the activities of enterprise innovation and business process reengineering. To this end, EasInnova is based on four pillars: (i) a Model-Driven Engineering, with the adoption of a Model-Driven Architecture, with its three layers (CIM. PIM, PSM), in the progressive development of business process models; (ii) a simple scheme, organised in three stages: AsIs, Transformation, ToBe, to guide the activities of the innovation undertaking; (iii) the *EasInnova* matrix, obtained by the interbreeding of the two above dimensions, that guides the innovation undertaking; (iv) the Low Code technology that capable of using a set of models in the PSM layer to generate a running application.

A preliminary test of *EasInnova* has been carried out with a simple case in the domain of home delivery food, as illustrated in the paper. A more extensive validation test is currently going on in an R&D department of an electro-medical company.